# PASS: Perturbation augmented space group structure sampling for transferable Fe-O machine learning interatomic potential

Zixiong Wei[1], Fei Shuang[1, 2], Poulumi Dey[1*]

[1] Department of Materials Science and Engineering, Faculty of Mechanical Engineering, Delft University of Technology, Mekelweg 2, 2628 CD Delft, The Netherlands

[2] State Key Laboratory of Nonlinear Mechanics, Institute of Mechanics, Chinese Academy of Sciences, Beijing 100190, China

## Abstract:

**Iron (Fe) oxidation is a fundamental phenomenon that is scientifically vital and technologically significant. Despite substantial experimental efforts, accurate atomistic modelling of Fe oxidation remains elusive, which requires a reliable interatomic potential, whose development hinges on a representative first-principles training dataset. However, Fe-oxygen (O) system is known for its structural complexity, rendering the generation of high-quality first-principles dataset challenging. In this work, we propose the Perturbation Augmented Space group structure Sampling (PASS) method to generate extensive dataset for Fe-O system consisting of small-cell structures (no more than 10 atoms). Based on atomic cluster expansion (ACE) framework, we present a systematic approach to developing a transferable machine learning interatomic potential (MLIP). The accuracy and capability of the ACE potential are thoroughly validated across pure Fe and Fe-O systems through bulk, surface, and interface properties. The large-scale Fe oxidation simulations with the ACE potential capture key features of the oxidation process, including the formation of an FeO-like structure consistent with recent experimental observations. This work demonstrates that the PASS method yields an accurate and transferable MLIP that can capture the reactive complexity of oxide growth while remaining computationally practical for extended systems.**

* Corresponding author: P.Dey@tudelft.nl (P. Dey)

# 1. Introduction

For decades, iron (Fe) oxidation has been a subject of substantial research effort due to its scientific significance and technological importance[1]. Scientifically, studying it not only helps understand Earth's geochemical evolution[2] but also provides insights into biological mechanisms, for example ferritins detoxification by catalyzing the oxidation of $Fe^{2+}$ ions[3]. Technologically, as energy production can be obtained through the heat of Fe powders oxidation, it potentially enables a safe, stable and high-density energy storage technology[4]. Moreover, it is usually linked with detrimental corrosion, whose prevention is key to achieve sustainability and durability of materials[5,6]. Therefore, Fe oxidation has been extensively investigated by all kinds of experimental techniques, such as low-energy electron diffraction[7,8], X-ray scattering[9], and *in situ* environmental scanning/transmission electron microscopy[10].

From an atomistic perspective, oxidation involves a series of critical atomic-scale events[11], including atomic adsorption, diffusion, and surface reconstruction. The combination of those crucial events poses a challenge to experimental studies, that is how to achieve three-dimensional atomic resolution for structural and chemical information *in situ*[12]. On the other hand, this challenge can be better tackled by atomistic modelling, especially molecular dynamics (MD) simulation, where structural and chemical information at the atomistic level, and thermodynamics and kinetics information can be obtained simultaneously[13]. Nevertheless, accurately modelling Fe oxidation is challenging due to the complexity of Fe-oxygen (O) system. First, multiple Fe oxides exist under different conditions[14,15], with wüstite (FeO), hematite ($Fe_2O_3$), and magnetite ($Fe_3O_4$) being the most stable ones at different O concentrations. Furthermore, various crystal structures and different stoichiometries add up to the structural complexity. For instance, $Fe_2O_3$ has a corundum structure, while FeO has a rock-salt structure with non-stoichiometric characteristics because of its high vacancy defect concentration and complex arrangements of point defect clusters[16]. Consequently, the concurrent occurrence of multiple kinetic events in oxidation, together with oxides' structural complexity, renders a reliable, robust and transferable interatomic potential imperative. Not only should it be capable of capturing the elementary reaction steps during initial oxidation but also of distinguishing between the nucleation, growth and phase transformations among the various iron oxide phases during the oxide layer growth.

Notably, reactive force field (ReaxFF), because of the explicit usage of bond order in its functional form, is effective in describing bond breaking and forming processes. This makes it quite suitable for the simulation of oxidation process[17]. Several ReaxFFs had been developed for the Fe-O system[18,19] and applied to study the oxidation process[9,20]. However, after comparing six ReaxFFs on the prediction of liquid Fe oxide properties, Thijs et al. found significant differences of properties obtained by them[21]. We previously benchmarked three ReaxFFs on their performance for O adsorption on Fe surfaces, which is the first yet quite important step in oxidation, but found that they had unsatisfactory descriptive capability, as the accuracy of the adsorption energy with regard to first-principles calculation is insufficient[22]. Therefore, a more accurate and broadly transferrable Fe-O interatomic potential is needed.

For the past two decades, owing to the capability of learning from large data and the achieved near-quantum accuracy, machine learning interatomic potential (MLIP) has attracted lots of attention[23,24]. Compared to empirical interatomic potentials, MLIPs do not have a pre-defined functional form. Instead, they use atomic descriptors (also called structural fingerprints) to encode the local atomic environments (LAEs). Various atomic descriptors have emerged to capture the essential characteristics of LAEs, including atom-centered symmetry functions[25], smooth overlap of atomic positions (SOAP)[26], moment tensor[27], and atomic cluster expansion (ACE)[28]. Despite the different varieties, the general development workflow is the same. After high-fidelity first-principles dataset is obtained, the LAEs are first encoded to be descriptors which are subsequently used as input to ML algorithm, while the configurations' corresponding energies, forces, and stresses act as output to be reproduced. In this way, an MLIP learns the mapping from training dataset LAEs to potential energy surfaces (PES). Therefore, the systematic generation of representative training dataset is essential to get an accurate and broadly transferrable MLIP, which should cover a large and relevant configurational space, especially for complex reactive systems.

Conventionally, training dataset is manually generated based on domain expertise, aiming at constructing configurations that are expected to be sampled during MD simulations. For example, to train an MLIP for fracture simulation, Zhang et al. included highly strained configurations that may appear at crack tip according to anisotropic classical linear elastic theory[29]. Since crucial LAEs in oxidation encompass not only various stable oxides, but also strained, defective, interfacial, and non-equilibrium configurations, it is unfeasible to manually enumerate all the necessary configurations based on domain expertise. Notably, based on extensively exploring all possible space groups of crystal structures, Poul et al. proposed the Automated Small SYmmetric Structure Training (ASSYST) strategy for constructing systematic training dataset[30,31]. Its effectiveness was demonstrated by various tests on out-of-distribution structures (i.e., not explicitly considered in the training dataset). The strategy was first applied to elemental magnesium[30] and further extended to multicomponent alloys[31]. However, ASSYST method does not investigate the coverage of LAEs of generated structures. It is demonstrated by Qi et al. that featurization of the configuration space could be used to help develop reliable MLIPs[32].

In this work, to extend ASSYST method, we propose the Perturbation Augmented Space group structure Sampling (PASS) method for building compact and diverse MLIP training dataset. Following the PASS workflow, we generate extensive and representative Fe-O dataset, composed of structures with no more than 10 atoms. An ACE potential is developed and subsequently validated on various properties across pure Fe and Fe oxides to demonstrate its accuracy and transferability. By applying the ACE potential to study large-scale Fe oxidation, we showcase the formation of FeO-like oxide, matching well with recent experiment findings. This work applies the sampling of small-cell structure to the simulation of large-scale oxidation for the first time, demonstrating that PASS is systematic, effective, and capable of enabling accurate and transferable MLIPs for chemically complex reactive processes.

# 2. Methods

## 2.1. Dataset generation and selection

The pyiron integrated development environment (IDE) is adopted as the dataset generation platform[33,34]. Firstly, the initial space group structures are generated by the PyXtal package[35]. After the stepwise relaxation is complete, the rattle and stretch perturbations are applied to structures using the ASSYST[31] tool implemented in pyiron IDE. To encode the LAEs, the SOAP descriptor[26,36] is adopted, as implemented in the DScribe package[37,38]. The parameters $r_{\mathrm{cut}}$ = 7, $n_{\mathrm{max}}$ = 12, $l_{\mathrm{max}}$ = 4, and $\sigma$ = 0.3 are used for constructing the SOAP descriptor.

## 2.2. DFT calculations

Collinear spin-polarized DFT calculations are performed using VASP[39,40] at T = 0 K[41,42]. The interactions between the valence electrons and the ionic core are described using the projected augmented wave (PAW) method[43]. The generalized gradient approximation (GGA) in the Perdew-Burke-Ernzerhof (PBE) form is adopted for exchange-correlation energy[44]. The Brillouin zone is sampled by the $\Gamma$-centered $k$-point mesh, with a consistent $k$-point spacing of 0.2 $Å^{-1}$. The Gaussian method is used to treat fractional occupancies with the Fermi surface smearing width of 0.05 eV. A cutoff energy of 520 eV is chosen for plane-wave basis set. The convergence criteria of the energy for the electronic self-consistency loop and the force for the ionic relaxation loop are set to $10^{-6}$ eV and 0.01 eV/Å, respectively. The diffusion barriers are obtained by performing climbing image nudged elastic band (CINEB) calculations[45], where the intermediate images are linked by a spring constant of 5 eV/$Å^2$.

## 2.3. MLIP training and validation

The ACE potential is trained by employing the PACEMAKER package[46–48]. The Finnis-Sinclair-type non-linear embedding function is adopted for both Fe and O elements. The simplified spherical Bessel functions are used as the radial basis functions. We use 800 B-basis functions per element, leading to 4,448 parameters. A global cutoff of 6 Å is used for neighbor list construction. The force-to-energy weight ratio in the loss function is set to $10^{-2}$. Both L1- and L2-regularization coefficients are set to $10^{-8}$. The parameters are optimized by the Broyden-Fletcher-Goldfarb-Shanno (BFGS) algorithm for 4,000 steps. Most validation tests are conducted in a workflow based on Atomic Simulation Environment (ASE) package[49]. We use the FIRE algorithm[50] with a force convergence criterion of $5\times10^{-2}$ eV/Å in all calculations where structural optimization is needed.

## 2.4. MD simulations

All MD simulations are performed using Large-scale Atomic/Molecular Massively Parallel Simulator (LAMMPS)[51]. For the diffusivity of O atoms in Fe bulk, a 10×10×10 body-centered cubic (BCC) supercell containing 2,000 Fe atoms is created, with 20 O atoms randomly inserted. Periodic boundary conditions are applied in all three directions. The system is relaxed at each temperature for 10 ps, followed by another 500 ps of production run. Isothermal-Isobaric (NPT) ensemble and canonical ensemble (NVT) with a timestep of 0.5 fs are adopted in relaxation and production run, respectively. The diffusion coefficients are obtained by fitting the mean square displacement of O atoms to the Einstein relation. For the Fe surface oxidation simulation, BCC(100) surface is selected and a 12×12×24 supercell is created, amounting to 3,456 Fe atoms. Then, 400 $O_2$ molecules are randomly distributed on top of the surface. During the oxidation simulation, periodic boundary conditions are adopted in both X and Y directions, while fixed boundary condition is applied along the Z direction, with both the top and bottom simulation box being set to be reflective walls. The atoms in three bottom layers of Fe slab are fixed to mimic the surface condition. Before the oxidation, the $O_2$ molecules and Fe slab are relaxed at 300 K in the NVT ensemble for 100 ps, respectively, so as for $O_2$ molecules randomly distributed over the surface and the initial configuration to be reasonable. In the oxidation process, the timestep is set to 1 fs and the temperature of the entire system is regulated at a constant 873 K by Nosé-Hoover heat bath method[52,53] for 3 ns. Open Visualization Tool (OVITO)[54] is utilized for visualizations.

# 3. Results

## 3.1. The PASS method

A central bottleneck in MLIP development for complex reactive materials is representative dataset construction, which we address through the introduction of PASS. It allows an ACE potential trained on small-cell structures to transfer to: (1) large structures, such as surfaces and interfaces; (2) application-relevant events, for example atomic diffusion and defect migration; and (3) complex reaction processes, for instance large-scale Fe oxidation.

### 3.1.1. Dataset generation and sampling workflow

The essential ideas behind PASS are first generating a comprehensive configurational space and then selecting the most representative structures while simultaneously capturing both equilibrium and non-equilibrium configurations. Specifically, two consecutive steps are performed, namely dataset generation and descriptor-based selection, as shown in Figs. 1(a) and (b), respectively.

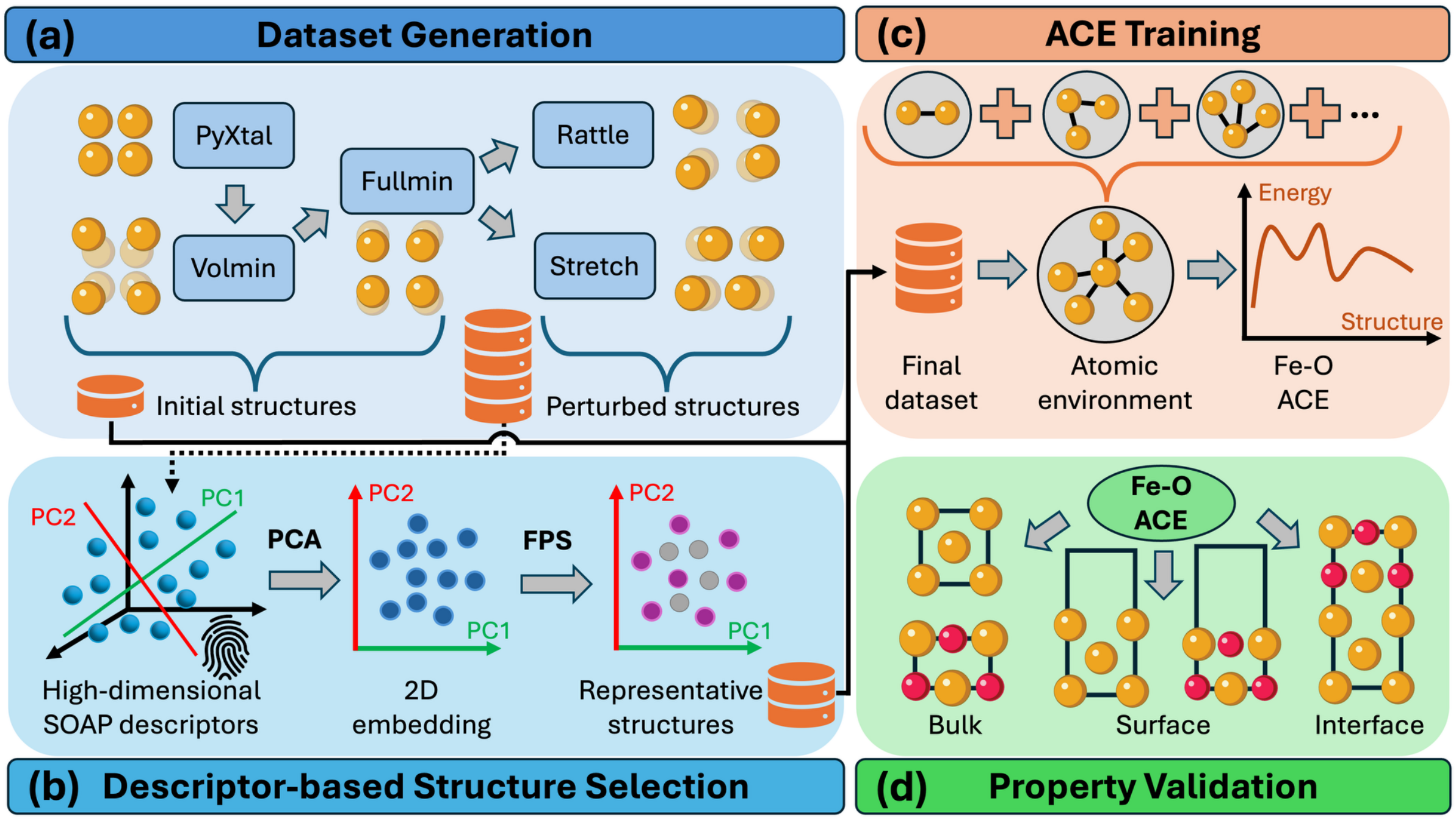

**Fig. 1. Overview of the PASS workflow for developing a transferable ACE potential for Fe-O system.** (**a**) Dataset generation. Space group structures are first generated using PyXtal and then optimized through stepwise volume and full structural relaxation to obtain initial equilibrium configurations. Perturbed structures are subsequently generated from the relaxed structures through systematic rattle and stretch operations, thereby introducing non-equilibrium LAEs. (**b**) Descriptor-based structure selection. The generated perturbed structures are encoded using high-dimensional SOAP descriptors, projected into a 2D embedding space by PCA, and selected by FPS to retain the most representative structures while reducing redundancy. (**c**) ACE training. The DFT-labelled final dataset is used to train an Fe-O ACE potential. (**d**) Property validation. The trained ACE potential is further systematically validated on bulk, surface, and interface properties across pure Fe and Fe-O systems to assess its accuracy and transferability.

Two parts of dataset, namely initial and perturbed structures, are constructed in the dataset generation step. The initial structures are obtained by stepwise relaxation of space group structures, following the protocols proposed in ASSYST method by Poul et al.[30,31], and the perturbed structures are constructed by applying perturbations (i.e., rattle and stretch) to the relaxed initial structures. As shown in Fig. 1(a), PyXtal generates structures according to user-specified space groups with the maximum number of atoms and atomic compositions provided as input parameters.

In this work, we set 10 to be maximum number of atoms per structure for both Fe and Fe-O systems. Subsequently, the volumes of PyXtal structures are relaxed, leading to Volmin structures. Afterwards, Volmin structures are fully relaxed to form Fullmin structures. Note that for the purpose of obtaining equilibrium configurations in a less computational demanding way, the stepwise relaxation processes are performed at a lower density-functional theory (DFT) accuracy in terms of both energy convergence criterion ($10^{-4}$ eV) and $k$-point density (0.5 $Å^{-1}$). Finally, PyXtal, Volmin, and Fullmin structures are collected, forming the collection of initial structures for the final high-fidelity DFT labelling. To generate perturbed structures, two separate actions (rattle and stretch) are taken on Fullmin structures. As proposed originally[31], the rattle is applied by changing atomic positions by normally distributed noise with a standard deviation of $\sigma_{\mathrm{rattle}}$ and then repeated $n_{\mathrm{rattle}}$ times. A uniformly random strain matrix is coupled with random displacements in this action. The stretch is implemented $n_{\mathrm{stretch}}$ times by changing cells using diagonal and off-diagonal entries. Two strain modes are randomly selected i.e., 70% hydrostatic (diagonal) and 30% shear (off-diagonal), with strain magnitudes $\epsilon_{\mathrm{hydro}}$ and $\epsilon_{\mathrm{shear}}$, respectively.

The first crucial element in PASS method is to extensively sample the configurational space in a controlled way. Specifically, a grid of perturbation parameters (i.e., $\sigma_{\mathrm{rattle}}$, $\epsilon_{\mathrm{hydro}}$, and $\epsilon_{\mathrm{shear}}$) is applied to significantly increase the number of perturbed structures. The values of those parameters for both Fe and Fe-O systems are presented in Supplementary Table S1. To summarize, the space group structures are generated first, providing compact and diverse crystalline prototypes. Then, extensive perturbations are applied, aiming at generating non-equilibrium LAEs. In this way, a comprehensive configurational space is generated. The second crucial element in PASS method lies in the descriptor-based selection or sampling process, as shown in Fig. 1(b). After filtering out structures with unphysically short interatomic distances, high-dimensional SOAP descriptors are utilized to encode the perturbed structures' LAEs. Consequently, the structural characteristics can be systematically quantified, enabling a convenient comparison between LAE differences. Next, principal component analysis (PCA) is adopted to reduce the high-dimensional descriptor space to a 2D embedding space. Finally, farthest point sampling (FPS) method is used to select the most representative structures thus avoid redundancy. Combined with the initial structures (i.e., PyXtal, Volmin, and Fullmin), a structure pool is now established and will eventually be labelled with high-fidelity DFT to form the final training dataset. Note that PASS method is only applied to obtain Fe and Fe-O structures following the procedures described above. For the O part, we directly adopt structures from existing datasets, including OQMD[55,56] and OMat24[57]. The final training dataset consists of 42,063 structures in total, and its composition is shown in Supplementary Table S2.

### 3.1.2. The diversity and representativeness of the final dataset

The purpose of PASS method is not simply enlarging the training dataset but constructing a compact and representative one that spans a broad range of physically relevant LAEs. The initial space group structures provide chemically and crystallographically diverse equilibrium configurations, while the systematic and extensive rattle and stretch perturbations introduce non-equilibrium LAEs potentially associated with defective, strained, and transition state configurations. Then, descriptor-based selection removes redundancy by selecting structures that are maximally separated in the reduced SOAP descriptor embedding space. Therefore, the final dataset is designed to balance diversity and efficiency such that it retains representative LAEs without relying on manually designed oxidation pathways.

Figure 2 illustrates the diversity and representativeness of the final dataset for the Fe-O system generated following PASS method. As shown in Fig. 2(a), the distribution of the atomic energy with respect to nearest neighbor distance is displayed, along with O content and structure count. Due to the filtering process, the minimum atomic distances in Fe and Fe-O systems are 1.8 and 1.6 Å, respectively. This ensures that all configurations are physical, i.e., not overly distorted. In addition, the atomic energy distributes in a wide range, indicating that a broad region of the energy landscape is covered, which is formed by including both equilibrium and non-equilibrium configurations. Furthermore, data points with different O content are widely scattered, meaning that diverse Fe-Fe, Fe-O, and O-O coordination environments are included, indicative of quite distinctive configurations being generated. Then, the formation enthalpy of all structures as a function of the atomic percentage O content is shown in Fig. 2(b). A large compositional space is covered by the final dataset, including pure Fe, Fe-O systems, and O structures. To demonstrate that the small-cell PASS dataset is representative of larger and more complex structures in application-relevant events such as fracture and oxidation processes, it is essential to compare LAEs with those of existing datasets regarding various applications. It is reported that atomic descriptors enable both phase classification and detection of point defects and interfaces[62]. To this end, the SOAP descriptor is used to represent the LAEs of both Fe and Fe-O systems, followed by PCA performed for dimension reduction to ensure a clearer visual comparison. Figure 2(c) and (d) display the LAE comparison between PASS dataset and selected ones for Fe and Fe-O systems, respectively. Note that for pure Fe system each point stands for the LAE of the whole structure, while for Fe-O system each point is the LAE of individual atom.

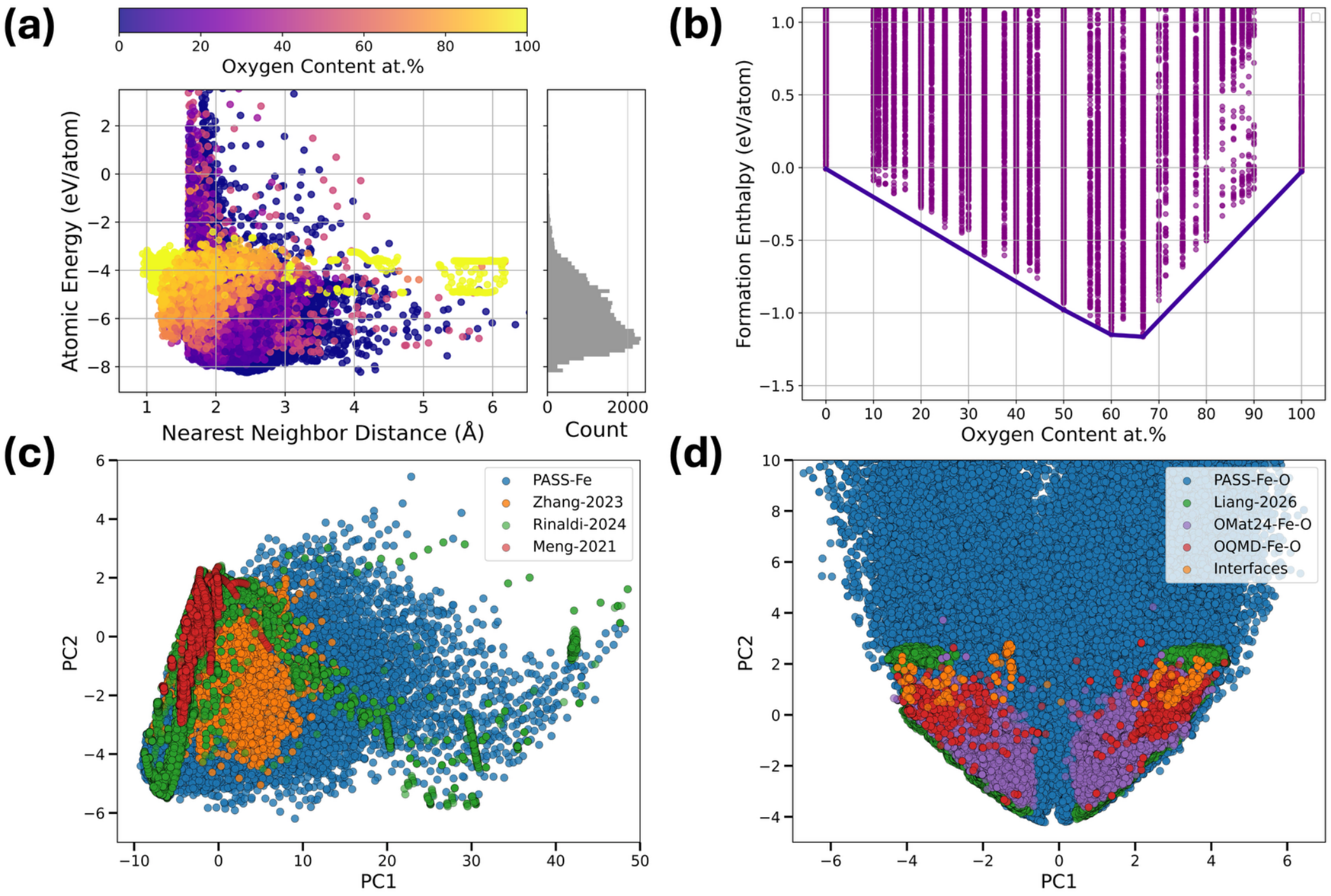

**Fig. 2. Diversity and representativeness of final dataset for Fe-O system.** (**a**) Distribution of atomic energy as a function of nearest neighbor distance, with data points colored by O content and accompanying histogram showing the structure count. (**b**) Formation enthalpy of the final dataset as a function of O atomic percentage. (**c**) PCA projection of SOAP descriptors for pure Fe, compared with existing Fe datasets including Meng-2021[58], Rinaldi-2024[59], and Zhang-2023[29]. Each point represents one structure-level SOAP average. (**d**) PCA projection of SOAP descriptors for Fe-O, compared with existing Fe-O datasets (i.e., Liang-2026[60] , OMAT24 Fe-O[57], and OQMD Fe-O[55,56]) and interfaces structures[61]. Each point represents the LAE of an individual atom.

For the Fe system, three datasets are chosen for comparison, i.e., Meng-2021[58], Rinaldi-2024[59], and Zhang-2023[29]. Meng et al. developed a neural network interatomic potential for the Fe-H system[58], from whose training dataset we extracted the configurations containing only Fe and mark them as Meng-2021. A large variety of configurations is included in Meng-2021, consisting of self-interstitial atoms, vacancies, surfaces, and dislocations. To comprehensively sample both atomic positions and magnetic moments, Rinaldi et al. generated extensive Fe configurations associated with various properties (e.g., elastic constants, phonon, and transformation path)[59], which we mark as Rinaldi-2024. Based on anisotropic classical linear elastic theory, Zhang et al. generated structures representative of the large deformations at crack tip[29], here marked as Zhang-2023. As can be seen in Fig. 2(c), Rinaldi-2024 has more structural diversity than Meng-2021, while Zhang-2023 occupies a large configurational space uncovered by the others, especially corresponding to the highly strained crack-tip configurations. Compared to the three datasets, the LAEs in our dataset (marked as PASS-Fe) could represent those complex configurations (e.g., vacancies and dislocations) in various applications (e.g., fracture and tensile stretch). Furthermore, LAEs in PASS-Fe dataset are even more widespread and diverse, which may be related to more complex structures such as transition states. The comparison not only demonstrates small-cell PASS dataset's outstanding representativeness but also shows its structural complexity.

For the Fe-O system (see Fig. 2(d)), three other datasets along with several interface structures are included for comparison. Based on previous training dataset focusing on the defective FeO structures as well as various FeO surfaces[16], Liang et al. included various $Fe_3O_4$ structures and employed deep learning methods to train an Fe-O interatomic potential[60], referred to as Liang-2026. In addition, Fe-O structures are extracted from both OQMD[55,56] and OMat24[57] datasets. OQMD contains equilibrium or near-equilibrium commonly occurring Fe-O crystal structures, while OMat24 consists of diverse non-equilibrium atomic configurations. Including those two datasets enables a more general LAE comparison beyond that of defective FeO and $Fe_3O_4$ dataset of Liang et al. Moreover, since the interfacial transformations among Fe, FeO and $Fe_3O_4$ are essential to the Fe oxidation process[63], complex interfaces[61] between Fe and $Fe_3O_4$ are also included. As can be seen in Fig. 2(d), OMat24 Fe-O covers more space than OQMD Fe-O, indicating its inclusion of non-equilibrium configurations. The PASS-Fe-O dataset not just cover the configurational space of all other datasets but also extend beyond it, owing to the extensive perturbations. This clearly demonstrates the representativeness and diversity of the final dataset. The separate comparison of PASS-Fe-O with each dataset is shown in Supplementary Fig. S1.

## 3.2. Systematic validation of the ACE potential

As shown in Fig. 1(c), the ACE framework is adopted for MLIP training. It should be noted that we do not explicitly include magnetic degrees of freedom in the training process. To simplify, we use ferromagnetic (FM) ordering for Fe element in all DFT calculations for labelling the final dataset. The parameterized ACE potential yields the training mean absolute errors (MAEs) of 34.59 meV/atom and 123.93 meV/Å for energy and force, respectively. 5% of all structures is randomly selected as test set and the testing MAEs are 36.15 meV/atom and 132.76 meV/Å for energy and force, respectively. As schematically illustrated in Fig. 1(d), to verify the accuracy and demonstrate the transferability of the ACE potential, a series of validation tests are subsequently performed in a progressive manner, starting from pure Fe to Fe-O system and from thermodynamics to kinetics, covering structural, mechanical, and chemical properties.

### 3.2.1. From structural to mechanical properties for pure Fe system

We start the validation by comparing structural properties of pure Fe. Figure 3(a) shows the comparison of the equation of states (EOS) for both BCC and face-centered cubic (FCC) Fe phases obtained by DFT and the ACE potential. Note that FM and antiferromagnetic (AFM) orderings are considered in DFT calculations for BCC and FCC Fe phases, respectively. A good agreement can be found for both phases, except for the large volumes of FCC Fe where a slight deviation is observed. The inset in Fig. 3(a) displays a good match between ACE predications and DFT results around the equilibrium area of BCC Fe EOS curve. The equilibrium lattice constants at finite temperature obtained by ACE potential are shown in Fig. 3(b), in comparison with experimental data[64]. The ACE potential can well capture the influence of temperature and accurately describe the lattice thermal expansion of both phases. The Bain path model is widely used to study the martensitic transformation in steel[65], in which BCC Fe changes into FCC Fe (or vice versa) via a continuous tetragonal distortion (i.e., changing the c/a ratio). The comparison between atomic energy with respect to the varied c/a ratio predicted by the ACE potential and calculated by DFT is shown in Fig. 3(c). Since FM ordering is considered when labeling the final dataset, only FM ordering is applied in DFT calculations for comparison. Even without explicit inclusion of relevant structures, the ACE potential can correctly describe the variations of potential energy in the Bain path, especially the positions of local minimum/maximum, illustrating that the structural component of the phase transformation is captured.

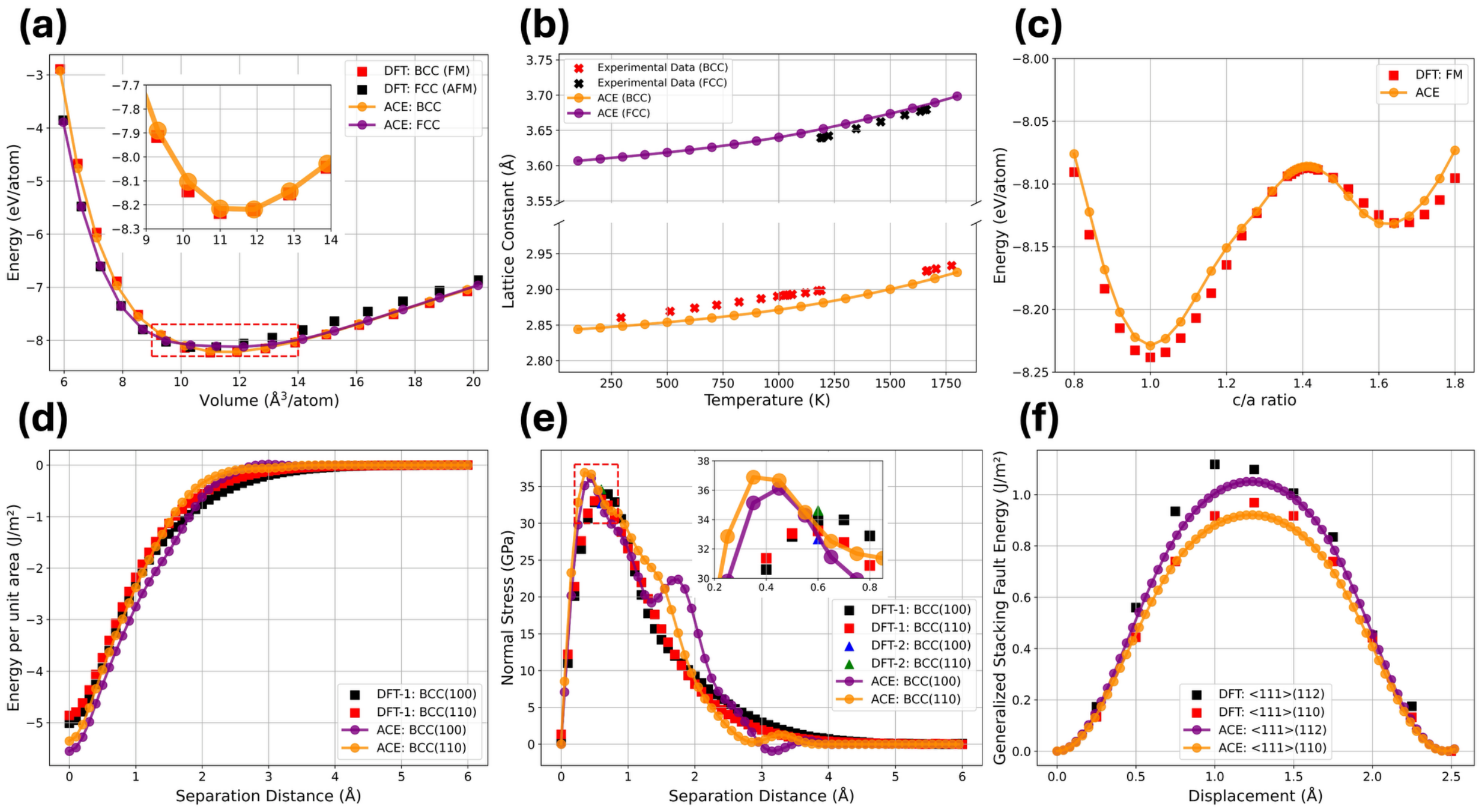

**Fig. 3. Validation of the ACE potential for structural and mechanical properties of pure Fe.** (**a**) EOS curves for BCC and FCC Fe predicted by the ACE potential and compared with DFT reference data. FM and AFM orderings are used in the corresponding DFT calculations for BCC and FCC Fe, respectively. (**b**) Temperature-dependent equilibrium lattice constants of BCC and FCC Fe predicted by the ACE potential and compared with experimental reference data. The experimental reference data is taken from Ref. [64]. (**c**) Bain path energy profiles as a function of the c/a ratio predicted by the ACE potential and compared with DFT reference data (FM setting). (**d**) Rigid body separation energy curves for BCC(100) and BCC(110) planes predicted by the ACE potential and compared with DFT reference data. The DFT reference data is taken from Ref. [66]. (**e**) T-S curves obtained from the derivative of the energy-separation curves predicted by the ACE potential and compared with DFT reference data. The DFT reference data is taken from Ref. [66] (DFT-1) and Ref. [29] (DFT-2). (**f**) GSF energy profiles for BCC(110) and BCC(112) planes along the <111> slip direction predicted by the ACE potential and compared with DFT reference data. The DFT reference data is taken from Ref. [68].

Next, we validate the ACE potential's performance on mechanical properties. Brittle fracture is the dominant failure mechanism for BCC Fe, which happens in the form of cleavage[29]. As cleavage occurs on the plane with maximum normal stress, it is necessary to evaluate the normal stress. To this end, the rigid body separation is performed for BCC(100) and BCC(110) planes, and the resulting energy-separation curves are shown in Fig. 3(d). DFT reference data taken from Ref. [66] is added to enable a clear comparison. The universal binding energy relation of BCC(110) plane aligns well with DFT results, while for BCC(100) plane slight deviation from DFT data is observed. BCC(100) and BCC(110) surface energies are calculated from Fig. 3(d) to be 2.80 and 2.66 J/m$^2$, respectively, matching with DFT obtained values of 2.543[67] and 2.495[67] J/m$^2$. As shown in Fig. 3(e), the traction-separation (T-S) curves are then obtained by taking the derivative of the energy-separation curves[68]. The general trends match with DFT reference data[66] (i.e., DFT-1) and the ACE potential can predict distinct maximum normal stresses (see inset in Fig. 3(e)), which for BCC(100) and BCC(110) planes are found to be close to the DFT results of 32.68[29] and 34.62[29] GPa (i.e., DFT-2, marked by blue and green triangles, respectively). Although local minima are present, the T-S curves are quite smooth compared to those of benchmarked MLIPs in Ref. [68]. Moreover, since the generalized stacking fault (GSF) energy is crucial for understanding nanoscale plasticity, ideal shear strength, and how crystalline materials deform[69,70], it is beneficial to add GSF profiles in the validation. Figure 3(f) shows the GSF energy of BCC(110) and BCC(112) planes slipping along <111> direction as predicted by the ACE potential, in comparison with DFT reference data[68]. The ACE potential can distinguish the shear deformations on the two planes, obtaining unstable (i.e., maximum) stacking fault energies close to the DFT results.

Subsequently, we validate the ACE potential's capability of describing various defects. Table 1 presents the calculated vacancy, self-interstitial atoms (SIA), and surface formation energies and the corresponding reference data. Two MLIPs specially trained for pure Fe, namely Gaussian approximation potential (GAP)[67] and moment tensor potential (MTP)[71], are selected along with DFT reference data. The ACE potential distinguishes the energetics of various defects, qualitatively matching with DFT, GAP[67] and MTP[71] results. Remarkably, none of the relevant configurations is explicitly included in the training dataset, thereby demonstrating the transferability of the ACE potential.

**Table 1. Validation of the ACE potential for point defects and surface formation energetics in pure Fe, in comparison with DFT and two MLIPs for pure Fe (i.e., GAP[67] and MTP[71]). "NN" represents nearest neighbor, "2NN" refers to the second nearest neighbor, and beyond. "T" and "O" stand for tetrahedral and octahedral, respectively. The notation "< >" specifies the exact geometric orientation and configuration of the SIA defect.**

| Property | ACE | DFT | GAP[67] | MTP[71] |
|---|---|---|---|---|
| Vacancy formation energy (eV) | | | | |
| mono-vacancy | 1.98 | 2.22[67], 2.184[71] | 2.26 | 2.161 |
| di-vacancy 1NN | 3.96 | 4.24[67], 4.231[71] | 4.41 | 4.124 |
| di-vacancy 2NN | 3.85 | 4.20[67], 4.161[71] | 4.30 | 4.063 |
| di-vacancy 3NN | 4.00 | 4.45[67], 4.396[71] | 4.55 | 4.308 |
| di-vacancy 4NN | 3.90 | – | 4.48 | – |
| di-vacancy 5NN | 3.97 | – | 4.47 | – |
| SIA formation energy (eV) | | | | |
| T site | 4.41 | 4.79[67], 4.548[71] | 4.75 | 4.87 |
| O site | 5.22 | 5.58[67], 5.333[71] | 5.53 | 5.267 |
| <100> | 4.96 | 5.48[67], 5.206[71] | 5.47 | 5.115 |
| <110> | 4.03 | 4.37[67], 4.094[71] | 4.21 | 4.282 |
| <111> | 4.51 | 5.13[67], 4.833[71] | 4.90 | 4.919 |
| Surface energy ($J/m^2$) | | | | |
| BCC(100) | 2.80 | 2.543[67], 2.525[71] | 2.547 | 2.543 |
| BCC(110) | 2.66 | 2.495[67], 2.524[71] | 2.499 | 2.543 |
| BCC(111) | 2.83 | 2.752[67], 2.714[71] | 2.756 | 2.723 |
| BCC(211) | 2.86 | 2.629[67] | 2.612 | – |

### 3.2.2. From thermodynamics to kinetics for Fe-O system

As the ACE potential is trained on small-cell PASS dataset, its transferability to larger systems needs to be validated. In addition, the accurate description of key oxidation events is significantly crucial to reliably simulate the process. Therefore, it is imperative to validate the ACE potential's capability of capturing the thermodynamics and kinetic properties of Fe-O system, as well as its transferability to complex structures and critical events.

**Table 2. Lattice constants (a for cubic lattice, a and c for hexagonal lattice), bulk modulus B, and formation enthalpy $\Delta H_f$ of Fe and representative Fe oxides predicted by our ACE potential, in comparison with DFT and ACE-Bienvenu[72]. DFT reference data is taken from Ref. [72]. The values in the parentheses are the results of DFT calculations with *U* correction. Lattice constants, bulk modulus, and formation enthalpy are in the units of Å, GPa, and eV/atom, respectively.**

| System | Property | ACE | DFT (+*U*)[72] | ACE-Bienvenu[72] |
|---|---|---|---|---|
| BCC Fe | a | 2.834 | 2.83 (2.95) | 2.84 |
| | B | 175 | 188 (123) | 165 |
| FeO | a | 4.296 | 4.26 (4.31) | 4.29 |
| | B | 167 | 195 (164) | 181 |
| | $\Delta H_f$ | −0.9118 | −0.91 (−1.43) | −0.95 |
| $Fe_2O_3$ | a | 4.956 | 5.02 (5.07) | 4.98 |
| | c | 13.408 | 13.90 (13.90) | 14.06 |
| | B | 179 | 172 (191) | 156 |
| | $\Delta H_f$ | −1.1958 | −1.30 (−1.81) | −1.30 |
| $Fe_3O_4$ | a | 8.252 | 8.40 (8.47) | 8.40 |
| | B | 163 | 172 (191) | 165 |
| | $\Delta H_f$ | −1.1816 | −1.29 (−1.75) | −1.29 |

First, our ACE potential is validated on the structural properties of common Fe oxides, namely FeO, $Fe_2O_3$, and $Fe_3O_4$. Bienvenu et al. trained an ACE potential on various phases present in the experimental Fe-O phase diagram, with magnetism explicitly considered in an Ising-like manner[72]. Their training dataset consists of prototype structures, relevant defects, grain boundaries, dislocation configurations, and ASSYST-generated structures. We add this ACE potential into comparison and hereafter refer to it as ACE-Bienvenu to distinguish with our ACE potential. As shown in Table 2, our ACE potential can describe the typical Fe oxides' structural properties close to DFT results, with performance comparable to ACE-Bienvenu. It is noteworthy mentioning that neither the structures nor the magnetism of the studied oxides (i.e. FeO, $Fe_2O_3$, and $Fe_3O_4$) are explicitly represented in the training dataset of our ACE potential, again demonstrating the effectiveness of PASS method. The formation enthalpies exhibit the same qualitative trend as the DFT results, indicative of our ACE potential capturing the thermodynamics.

The oxidation process begins with O adsorption on the Fe surface, followed by O dissolution into the bulk and subsequent diffusion. It is beneficial to incorporate these event-relevant configurations to evaluate whether the ACE potential accurately captures the thermodynamics in the critical events. Therefore, the DFT results for O adsorption on various BCC Fe surfaces at different coverages from our previous study are adopted[22]. In addition, we perform DFT calculations of O dissolution at both octahedral and tetrahedral sites in BCC Fe, together with the energy barrier along the diffusion path between those sites. The atomic energies of those configurations predicted by the ACE potential are compared to DFT reference data. As shown in Fig. 4(a), though no related configurations are included in the training dataset, the predictions of atomic energies by the ACE potential match well with the ones obtained by DFT, indicative of the captured thermodynamics in the oxidation-relevant events. Then, to validate the transferability to larger and complex structures, the geometry optimizations of the studied Fe oxides in both bulk and surface forms are performed by DFT (without $U$ correction), where their magnetic moments are set according to the respective ground state magnetism. Moreover, as the interfacial configurations are valuable in oxidation simulation and FeO is formed initially during oxidation[63], two interface structures between Fe and FeO (i.e., Fe(100)||FeO(100) and Fe(110)||FeO(111)) are added. All the relaxation trajectories are extracted, and the corresponding energies are predicted by the ACE potential. Figure 4(b) displays the atomic energies predicted by the ACE potential alongside DFT reference data, confirming that the ACE potential captures well the energetics of diverse configurations. Furthermore, as shown in Fig. 4(c), the diffusivity of O in the BCC Fe bulk is simulated and the resulting diffusion coefficients are compared with experimental study[73]. Though there are substantial magnitude discrepancies, a consistent trend can be observed in terms of the influence of temperature on diffusion coefficient. The activation energy is further calculated and the one obtained by our ACE potential is closer to experiment than the one by ACE-Bienvenu.

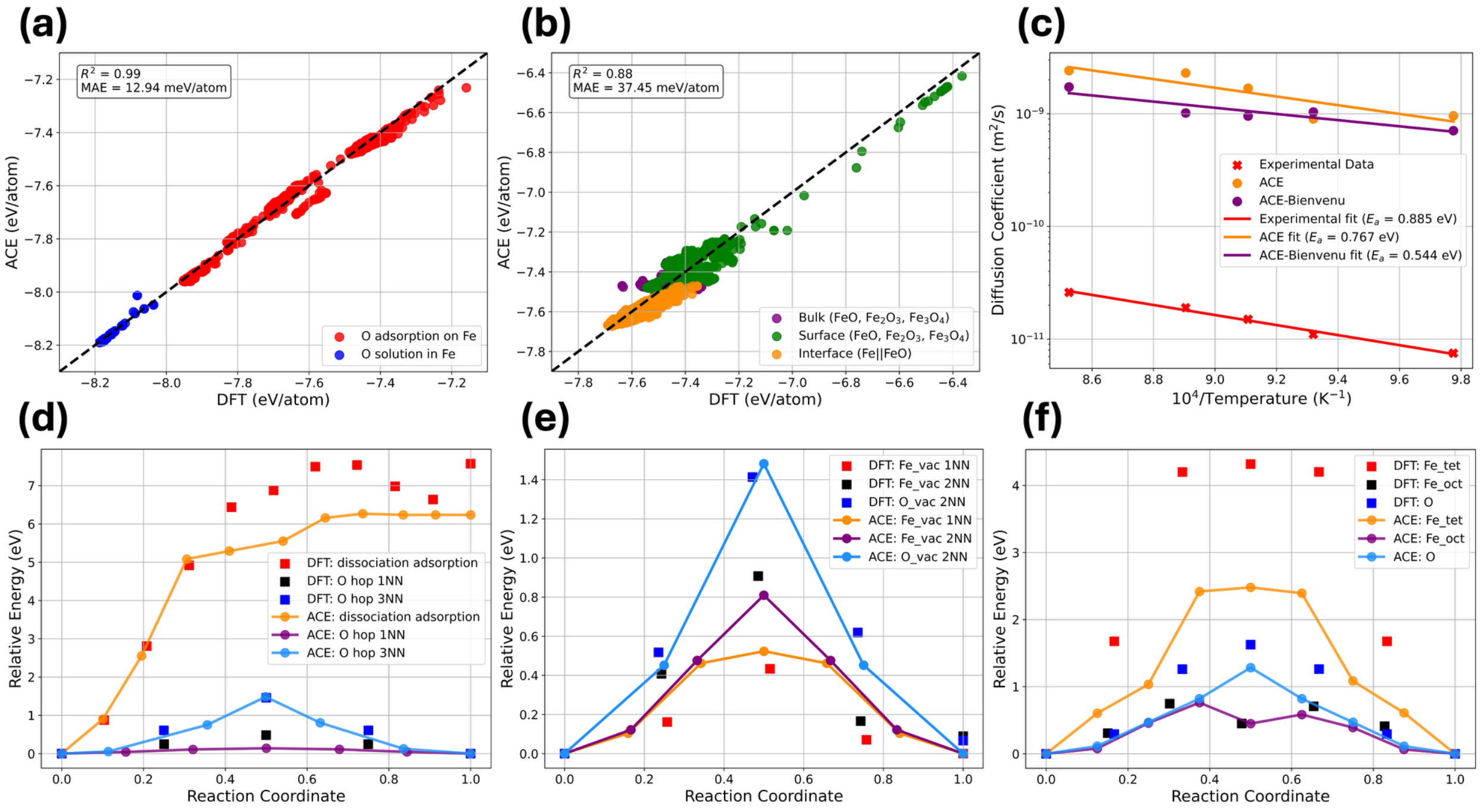

**Fig. 4. Validation of the ACE potential for thermodynamics and kinetic properties of Fe-O system.** (**a**) Comparison of atomic energies between ACE potential predictions and DFT calculations for oxidation-relevant configurations, including O adsorption on BCC Fe surfaces and O dissolution in BCC Fe. (**b**) Comparison of atomic energies between ACE potential predictions and DFT calculations for relaxation trajectories of Fe oxide bulks, surfaces, and Fe/FeO interface structures. (**c**) Temperature-dependent diffusion coefficients of O in BCC Fe predicted by our ACE potential, in comparison with ACE-Bienvenu and experimental data taken from Ref. [73]. (**d**) Energy barriers of $O_2$ dissociation adsorption on BCC Fe (100) surface and O hopping between octahedral sites along 1NN and 3NN pathways in BCC Fe predicted by the ACE potential, in comparison with DFT reference data. (**e**) Vacancy migration barriers in FeO obtained by the ACE potential, in comparison with DFT reference data. The vacancies migrate on the FeO(111) plane. (**f**) Vacancy migration barriers in $Fe_3O_4$ obtained by the ACE potential, in comparison with DFT reference data.

Next, to further validate the ability of the ACE potential to describe kinetic behaviors, the energy profiles of several oxidation-relevant processes are obtained from CINEB calculations. As shown in Fig. 4(d), $O_2$ dissociative adsorption on the BCC(100) surface and O hopping along 1NN and 3NN pathways in BCC Fe are considered, with the ACE predictions compared with DFT reference data. For the surface reaction, the ACE potential captures the overall increase in relative energy along the plotted reaction coordinate, showing good agreement with DFT in the initial part of the pathway. Nevertheless, deviations become more pronounced in the later part, where the relative energies are underestimated by the ACE potential. The initial and final configurations in the CINEB calculation are shown in Supplementary Fig. S2. For O diffusion in BCC Fe, the ACE potential correctly distinguishes the two hopping pathways, predicting a higher migration barrier for the 3NN hop than for the 1NN hop, consistent with DFT. In particular, the 3NN migration barrier is approximately 1.5 eV for both ACE and DFT, with the overall shape of the energy profile also reproduced. In contrast, the 1NN migration barrier is underestimated, with values of approximately 0.15 and 0.48 eV obtained by ACE and DFT, respectively. These results demonstrate that the ACE potential captures the relative ordering of the examined O migration barriers, although its quantitative accuracy remains pathway dependent. Together with the preceding thermodynamics validation, this comparison extends the assessment to non-equilibrium configurations associated with O surface reactions and diffusion in Fe.

The vacancy migration barriers in FeO and $Fe_3O_4$ are further evaluated below. First, vacancy migration on the FeO(111) plane is considered, including the 1NN and 2NN Fe vacancy migration paths and the 2NN O vacancy migration path. As shown in Fig. 4(e), the ACE potential reproduces the relative ordering of the three migration barriers. The 1NN Fe vacancy migration has the lowest barrier, while the 2NN O vacancy migration has the highest one. The ACE barriers are 0.52, 0.81, and 1.48 eV for the three paths, respectively, compared with 0.43, 0.91, and 1.41 eV obtained by DFT. Although differences remain along individual reaction paths, the comparison indicates that the ACE potential can distinguish the different vacancy migration processes in FeO. Then, vacancy migration in $Fe_3O_4$ is evaluated for Fe at tetrahedral and octahedral sites and for O at octahedral site, as shown in Fig. 4(f). The ACE potential captures the relative ordering of the migration barriers, with the tetrahedral Fe path showing the highest barrier and the octahedral Fe path the lowest one. The octahedral Fe migration profile follows the DFT reference reasonably well, while the O migration barrier is 1.28 eV, compared with 1.63 eV from DFT. A larger deviation is observed for tetrahedral Fe migration, where the ACE barrier is 2.48 eV rather than the DFT value of 4.32 eV. Thus, while the ACE potential captures the qualitative differences between these migration processes, further refinement of the tetrahedral Fe migration barrier could improve the quantitative description of oxide kinetics.

## 3.3. Application to large-scale oxidation simulation

After systematically validating the ACE potential on structural, mechanical, and oxidation-relevant properties, we further assess its capability of describing chemically evolving Fe-O systems. To this end, the ACE potential is applied to Fe surface oxidation at 873 K for 3 ns. It is reported by Tu et al.[63] that Fe oxidation follows the sequence of: (1) the nucleation of an initial FeO layer, (2) the growth of FeO to a certain thickness, and (3) the FeO-to-Fe3O4 structural transition at the Fe/FeO interface. Thus, we adopt three initial configurations to mimic the three distinct stages, namely unoxidized Fe surface, thin FeO layer on top of Fe surface, and thick FeO layer on top of Fe surface. Figures 5-7 show the structural evolution starting from an initially unoxidized surface and two configurations already containing an oxide layer. The atomic configurations and layer-resolved average Fe coordination numbers are examined together to characterize changes in the surface, oxide layer, and underlying Fe substrate. Subsequently, the radial distribution functions (RDF) of the oxide layers are compared with an FeO reference as shown in Fig. 8.

Figures 5(a-d) displays the oxidation of the initially unoxidized Fe surface. An oxide layer develops near the surface during the first 1 ns, while the underlying Fe substrate remains largely ordered. The coordination analysis is performed using a cutoff radius of 2.65 Å. The corresponding coordination profiles are shown in Fig. 5(e-h). The profiles distinguish the relatively unchanged BCC Fe substrate from the evolving near-surface region. At 0 ns, the average Fe coordination number is close to 8 in the bulk and decreases near the free surface. After 1 ns, the near-surface region exhibits average Fe coordination numbers of approximately 6, separated from the BCC Fe substrate by a transition region with intermediate values. Between 1 and 3 ns, the overall coordination profile changes comparatively little, although local rearrangements remain visible near the surface and the metal/oxide interface. This indicates that the largest changes in the average local coordination occur during the first nanosecond of the displayed sequence. The near-surface region approximates coordination numbers of 6, indicative of the formation of FeO-like structure.

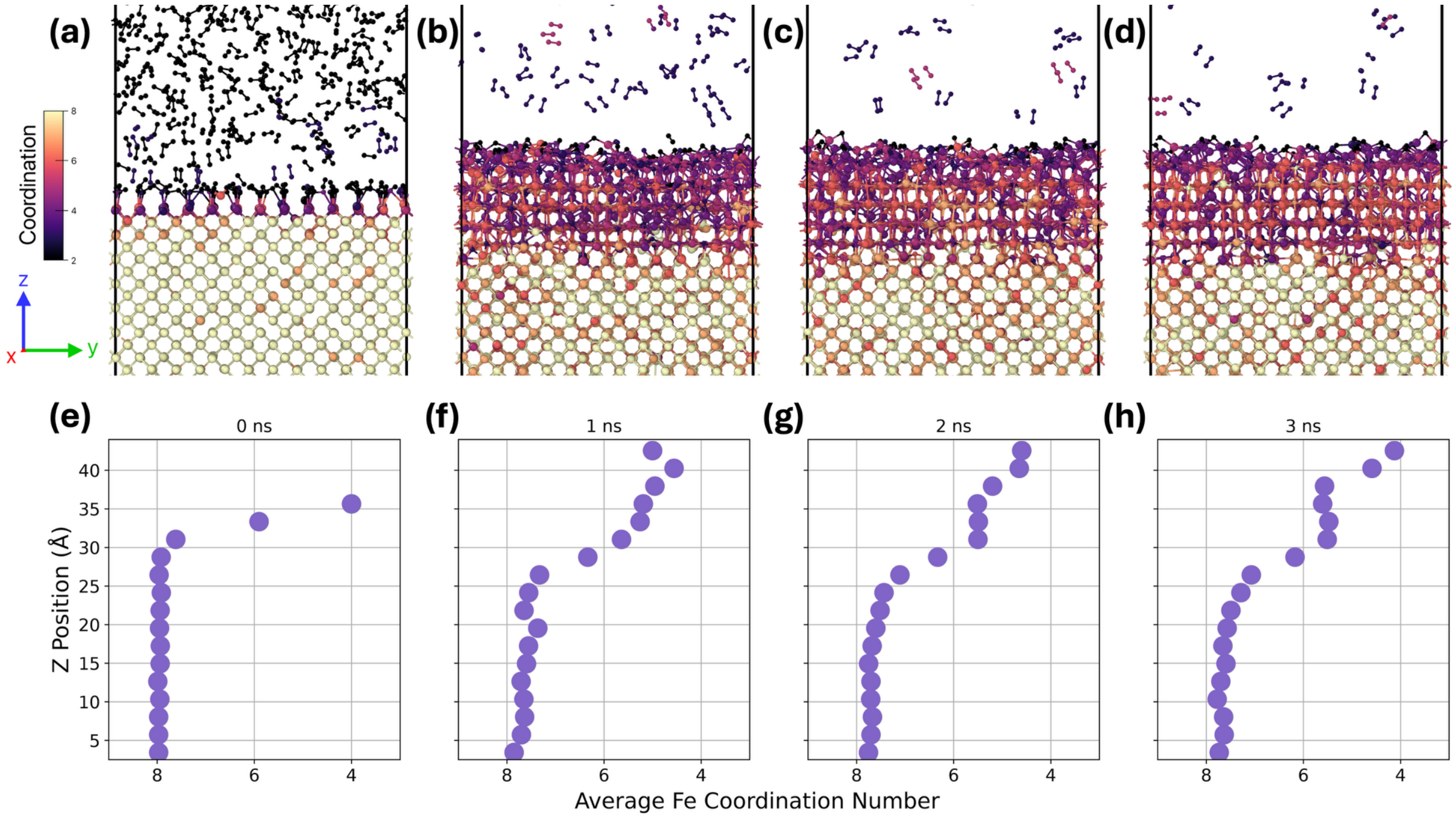

**Fig. 5. Oxidation of an initially unoxidized BCC Fe (100) surface using the ACE potential.** (**a-d**) Atomic configurations at 0, 1, 2, and 3 ns, respectively. Atoms are colored according to their local coordination numbers. (**e-h**) Corresponding layer-resolved average Fe coordination numbers along the Z direction.

Figure 6 shows the oxidation of a configuration in which a thin FeO layer is already present, which is to simulate the growth of FeO. Compared with the initial configuration (see Fig. 6(a)), the oxide-containing region (Figs. 6(b-d)) extends to higher Z positions after 1 ns, accompanied by changes in both the surface and interfacial atomic arrangements. During the subsequent evolution, the oxide layer becomes more structurally disordered, while the lower Fe substrate retains its ordered structure. As shown in Figs. 6(e-h), the average Fe coordination number remains close to 8 in the substrate and decreases to approximately 6 over much of the oxide-containing region, with lower values near the outer surface. The approximate coordination number of 6 in the oxide-containing region implies the growth of FeO-like structure. These results show that the ACE potential can describe continued structural evolution of FeO-like configuration in the presence of an existing oxide layer.

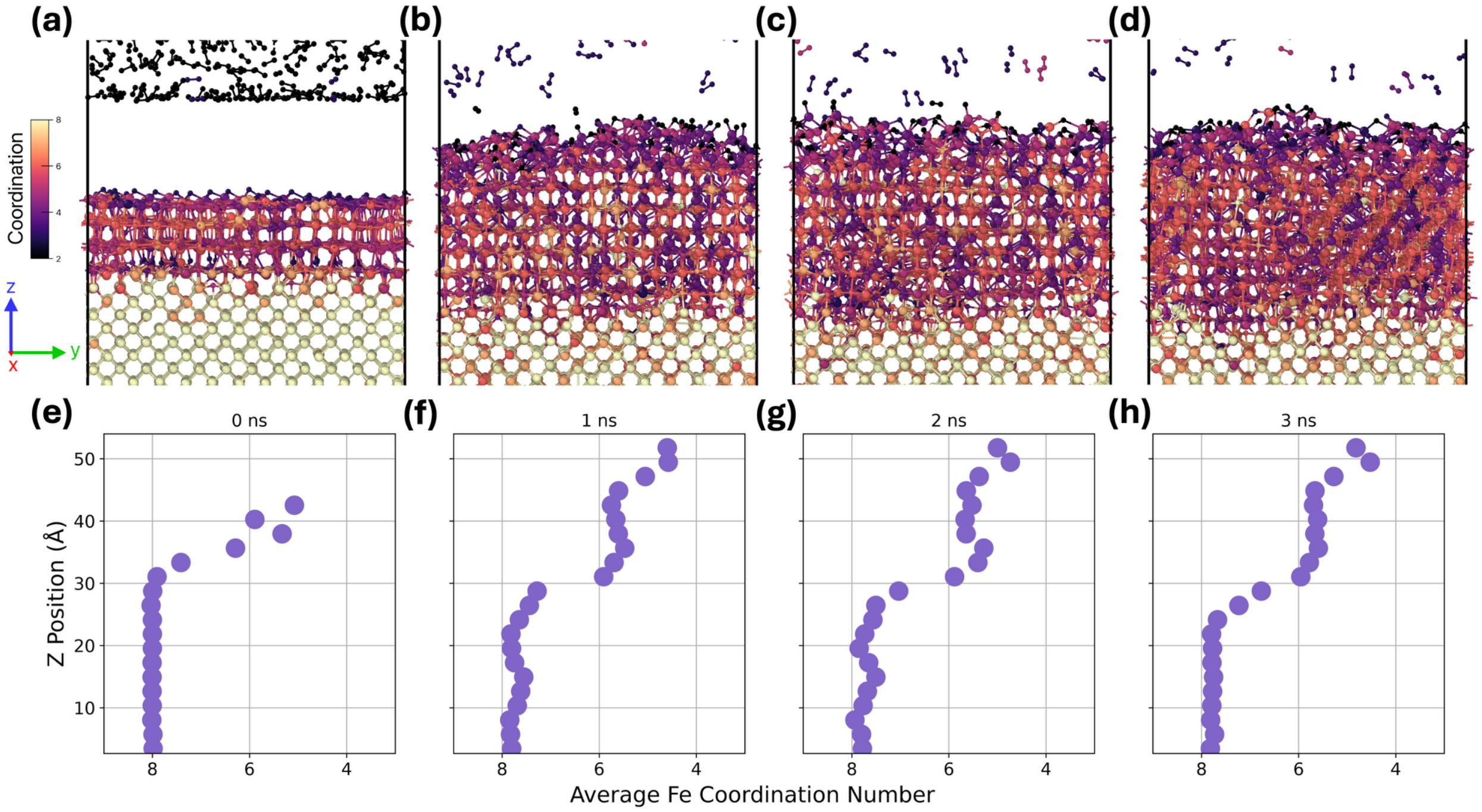

**Fig. 6. Oxidation of a thin FeO covered Fe surface using the ACE potential.** (**a-d**) Atomic configurations at 0, 1, 2, and 3 ns, respectively. The initial configuration contains a FeO layer with thickness of 6.4 Å above the BCC Fe substrate. Atoms are colored according to their local coordination numbers. (**e-h**) Corresponding layer-resolved average Fe coordination numbers along the Z direction.

Figure 7 shows the evolution of a configuration with a thicker and initially ordered FeO layer to simulate the transformation of FeO. The average Fe coordination number in the interior of the initial oxide is close to 6, in contrast to the value of approximately 8 in the underlying BCC Fe. During the simulation, the ordered oxide structure remains visible and extends to both directions, indicating the growth of FeO-like structure. Nevertheless, the surface and interfacial arrangements change, and more heterogeneous local environments develop within the oxide layer, indicative of the transformation of FeO. The coordination profiles in Figs. 7(e-h) retain a broad region with average Fe coordination numbers of approximately 5-6, connected to the BCC Fe substrate through a transition region. Compared with the initially unoxidized surface in Fig. 5, this sequence illustrates the structural evolution of a system in which an extended oxide layer is already present.

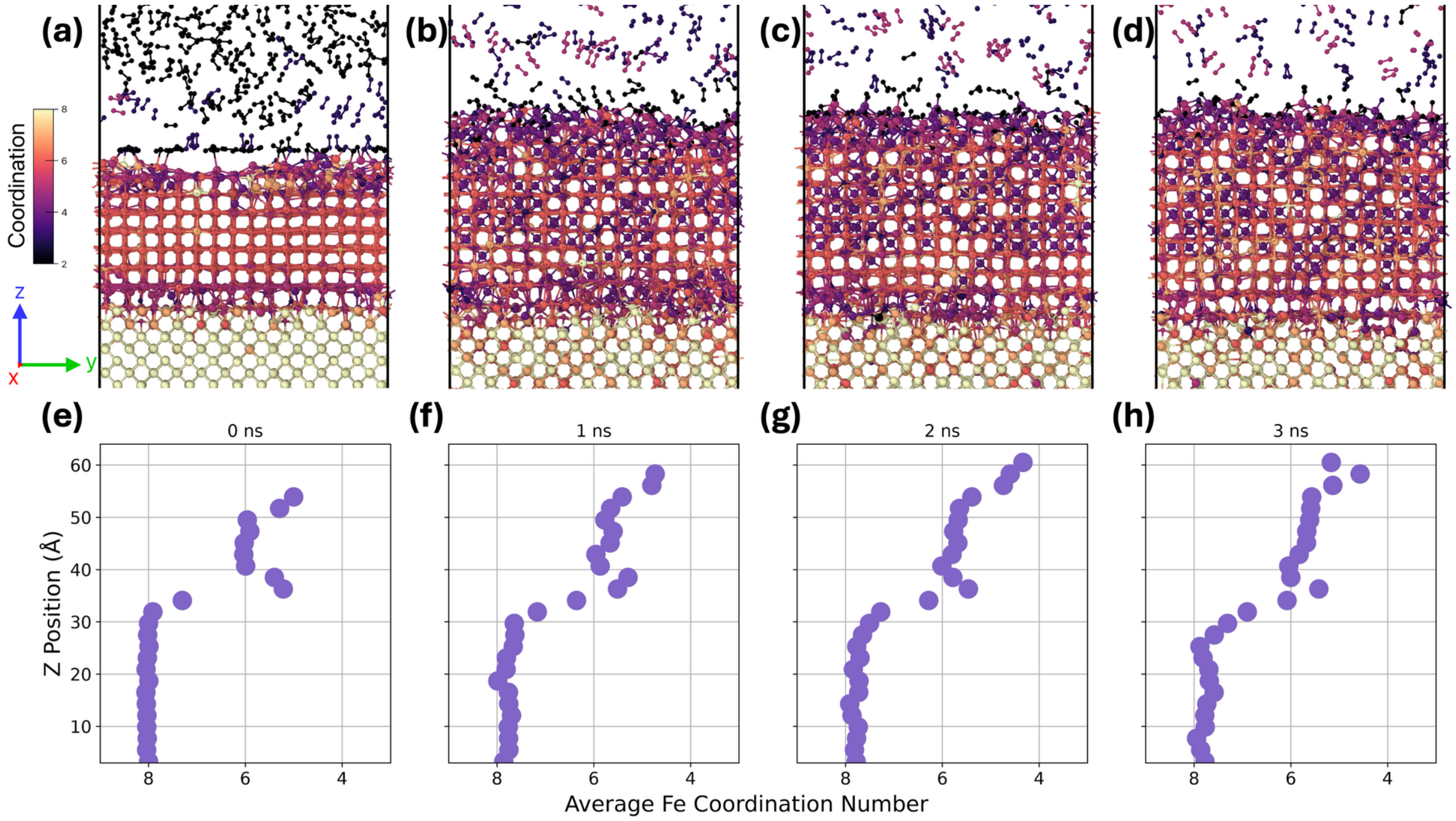

**Fig. 7. Oxidation of a thick FeO covered Fe surface using the ACE potential.** (**a-d**) Atomic configurations at 0, 1, 2, and 3 ns, respectively. The initial configuration contains a FeO layer with thickness of 16 Å above the BCC Fe substrate. Atoms are colored according to their local coordination numbers. (**e-h**) Corresponding layer-resolved average Fe coordination numbers along the Z direction.

To further probe the local structure of the oxide layers, the final oxide-containing regions from Figs. 5-7 are extracted and their pair RDFs are calculated. Figure 8(a) shows the DFT-relaxed FeO reference, while Figs. 8(b-d) shows the oxide layers obtained from the three oxidation sequences, respectively. In all three oxide layers, the first Fe-O peak is located near 2.0 Å, compared with approximately 2.1 Å in the FeO reference. The first prominent Fe-Fe and O-O peaks occur near 3 Å in both the reference and the simulated oxide layers. These similarities indicate that the oxide layers develop local pair correlations resembling those of FeO, although the peak positions do not coincide exactly. Compared with the FeO reference, the RDF peaks of the simulated oxide layers are broader, and several higher-shell features are less distinct. The peak shifts and broadening indicate that the oxidation products differ from the ordered reference structure. The comparison is therefore mainly based on the positions and overall features of the pair correlations rather than their peak heights. Together with the average Fe coordination numbers of approximately 5-6 in much of the oxide-containing regions, the RDF results support the presence of FeO-like short-range order. However, neither the coordination profiles nor the RDFs alone establish a unique oxide phase or its exact stoichiometry. Therefore, the formed layers are described as FeO-like rather than as perfect crystalline FeO.

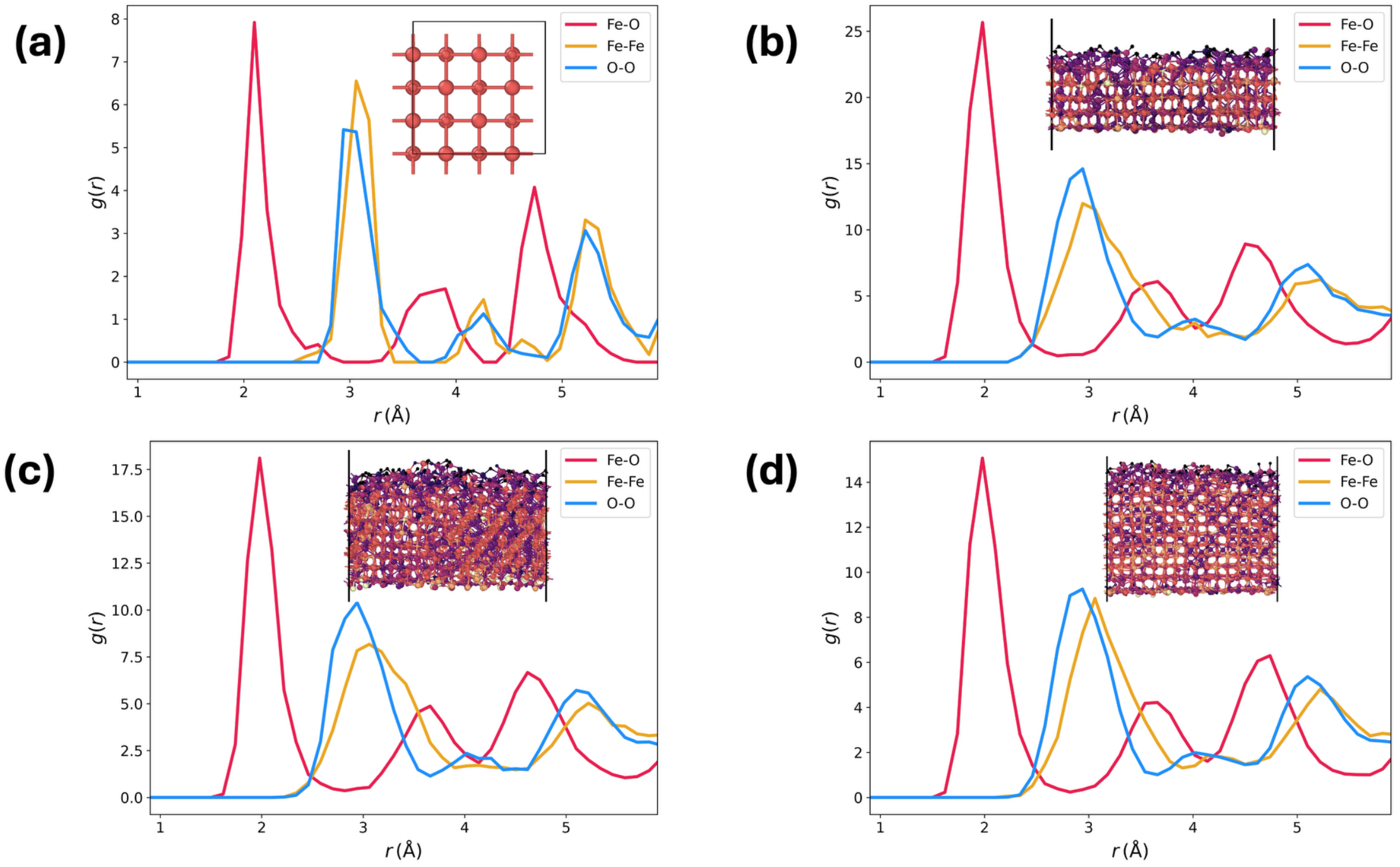

**Fig. 8. Pair RDF of the FeO reference and the oxide layers obtained from oxidation simulations.** (**a**) DFT-relaxed FeO. (**b-d**) Oxide layers extracted from final configurations at 3 ns in Figs. 5(d), 6(d), and 7(d), respectively. The Fe-O, Fe-Fe, and O-O pair correlations are shown as functions of interatomic distance. The insets display the corresponding structures.

To briefly summarize, the oxidation simulations show the formation and subsequent rearrangement of oxide layers, together with the evolution of the transition region above the BCC Fe substrate. The combined coordination and RDF analyses indicate FeO-like local ordering in the simulated oxide layers. This observation is qualitatively consistent with the initial FeO formation reported by Tu et al.[63], as discussed above. While the present analysis does not resolve subsequent phase transformations or establish a specific oxide-growth mechanism, the results demonstrate the potential of the small-cell-trained ACE potential to describe chemically evolving Fe–O configurations under the simulated conditions.

## 4. Discussion

We propose the PASS method, a symmetry-informed, perturbation-augmented, descriptor-based sampling strategy for building representative MLIP dataset of small-cell structures. The extensive validation tests and large-scale oxidation simulations demonstrate the effectiveness of PASS method. To the best of our knowledge, this work provides the first demonstration of a space group based small-cell structure sampling and training strategy for a transferable MLIP applied to large-scale and chemically complex Fe surface oxidation.

PASS avoids the heavy enumeration of manually generating dataset based on domain expertise. Additionally, training MLIP on the small-cell structures is advantageous, as it reduces the computational cost for the DFT calculations, thus allowing higher-fidelity DFT labelling. The most essential point is that MLIP learns the critical link from LAEs to PES, so small-cell structures' diversity can transfer to large and complex systems if the relevant LAEs are covered. The effectiveness of PASS can be attributed to the fact that it generates a broad range of LAEs, and the perturbation samples non-equilibrium configurations relevant to complex structures (e.g., defects and interfaces) and critical processes (e.g., adsorption and diffusion). Notably, the active learning strategy may be useful for targeted refinement, especially for rare oxidation events or highly defective interfaces, where uncertainty quantification plays a vital role[74].

Compared to the ASSYST strategy proposed by Poul et al.[31], PASS takes advantage of the expressive ability of atomic descriptors to represent LAEs. This makes it convenient to compare structural differences and further helps with the selection of the most representative structures. In summary, PASS not only makes sure that the MLIP is symmetry-informed by using space group structures, but also ensures that the broad range of LAEs are generated in a controlled manner by applying a mesh of perturbations and sampling representative structures based on atomic descriptors.

It should be noted that the current ACE potential does not explicitly consider magnetism and thus is not a spin-aware interatomic potential. Instead, it should be interpreted as a structurally transferable and computationally practical interatomic potential trained on a highly systematic, oxidation-agnostic, small-cell dataset under a fixed magnetic setting. Consequently, its applicability may be limited in regimes where magnetism strongly influences energetics and kinetics. For example, the FM to paramagnetic phase transition in BCC Fe depend on both spin and lattice degrees of freedom. In addition, the ACE potential cannot explicitly capture temperature-dependent spin fluctuations or magnetic disorder above the Curie or Néel temperatures. Its transferability comes from the extensive structural sampling and systematic validation, but future work could incorporate magnetic descriptors with explicit spin degrees of freedom, spin-labelled datasets, or separate magnetic treatment of oxidation states.

# 5. Conclusions

In this work, we extend the ASSYST-based small-cell structure generation strategy by combining systematic perturbations with descriptor-based selection and sampling in PASS method. The final dataset contains 42,063 structures, with the PASS-generated Fe and Fe-O configurations containing no more than 10 atoms. An ACE potential is trained on this dataset and yields test MAEs of 36.15 meV/atom and 132.76 meV/Å for energy and force, respectively. Systematic validation covers bulk structures, surfaces, interfaces, defects, and oxidation-relevant atomic processes. The ACE potential reproduces many of the tested structural and energetic trends, while deviations remain for specific kinetic properties, particularly tetrahedral Fe vacancy migration in $Fe_3O_4$.

The ACE potential is further applied to 3 ns Fe oxidation simulations starting from initially unoxidized and oxide-covered configurations. The simulations show the development and rearrangement of oxide layers above the BCC Fe substrate. The coordination profiles and pair RDFs indicate FeO-like short-range order, qualitatively consistent with the initial FeO formation discussed in the reported experimental work[63]. Together, these results highlight the transferability of LAEs generated from small-cell structures to the modelling of extended, chemically evolving Fe-O systems. The current ACE potential is limited by its fixed magnetic treatment, with scope for improving the quantitative prediction of certain migration barriers. Targeted refinement of the dataset and explicit treatment of magnetic degrees of freedom thus provide potential directions for further improvement.

# Data availability

Data will be made available upon reasonable request.

# Code availability

The files for the ACE potential will be publicly available once the manuscript is accepted.

# CRediT authorship contribution statement

**Zixiong Wei**: Conceptualization, Methodology, Software, Data curation, Formal analysis, Writing – original draft, Writing – review & editing. **Fei Shuang**: Conceptualization, Formal analysis, Writing – review & editing. **Poulumi Dey**: Funding acquisition, Resources, Project administration, Supervision, Writing – review & editing.

# Declaration of competing interest

The authors declare no competing interests.

# Acknowledgements

This research was carried out under project number T23003 in the framework of the Research Program of the Materials innovation institute (M2i) (www.m2i.nl) supported by the Dutch government. This work was supported by the Nederlandse Organisatie voor Wetenschappelijk Onderzoek (NWO; the Netherlands Organization for Scientific Research), Domain Science, for access to supercomputing facilities. We also acknowledge the use of the DelftBlue supercomputer provided by the Delft High Performance Computing Center (DHPC; https://www.tudelft.nl/dhpc). We thank Dr. Wanda Melfo Prada and Dr. Marga Zuijderwijk from Tata Steel Nederland Technology B.V. for fruitful discussions and insightful suggestions.

# Supplementary materials

Supplementary material associated with this article can be found, in the online version, at

# Supporting Information for

# PASS: Perturbation augmented space group structure sampling for transferable Fe-O machine learning interatomic potential

Zixiong Wei[1], Fei Shuang[1, 2], Poulumi Dey[1 1]

[1] Department of Materials Science and Engineering, Faculty of Mechanical Engineering, Delft University of Technology, Mekelweg 2, 2628 CD Delft, The Netherlands

[2] State Key Laboratory of Nonlinear Mechanics, Institute of Mechanics, Chinese Academy of Sciences, Beijing 100190, China

[1]: Corresponding author
E-mail address: P.Dey@tudelft.nl

## Supplementary Tables

**Table S1. The values of parameters for perturbating both Fe and Fe-O systems.**

| Action | System | Parameter | Value |
|---|---|---|---|
| Rattle | Fe | $\sigma_{\mathrm{rattle}}$ | 0.1, 0.2, 0.3, 0.4, 0.5 |
| | | $n_{\mathrm{rattle}}$ | 10 |
| | Fe-O | $\sigma_{\mathrm{rattle}}$ | 0.1, 0.2, 0.3, 0.4, 0.5 |
| | | $n_{\mathrm{rattle}}$ | 2 |
| Stretch | Fe | $\epsilon_{\mathrm{hydro}}$ | 0.05, 0.1, 0.2, 0.3, 0.4 |
| | | $\epsilon_{\mathrm{shear}}$ | 0.005, 0.01, 0.05, 0.1, 0.2 |
| | | $n_{\mathrm{stretch}}$ | 10 |
| | Fe-O | $\epsilon_{\mathrm{hydro}}$ | 0.05, 0.1, 0.2, 0.3, 0.4 |
| | | $\epsilon_{\mathrm{shear}}$ | 0.005, 0.01, 0.05, 0.1, 0.2 |
| | | $n_{\mathrm{stretch}}$ | 2 |

## Supplementary Tables

### Table S2. Composition of the final training dataset.

| System | Subset | Number of structures |
|---|---|---|
| Fe | Initial | 3,063 |
| | Perturbed | 10,000 |
| Fe-O | Initial | 7,774 |
| | Perturbed | 20,000 |
| O | OQMD | 33 |
| | OMat24 | 1,193 |
| Total | | 42,063 |

## Supplementary Figures

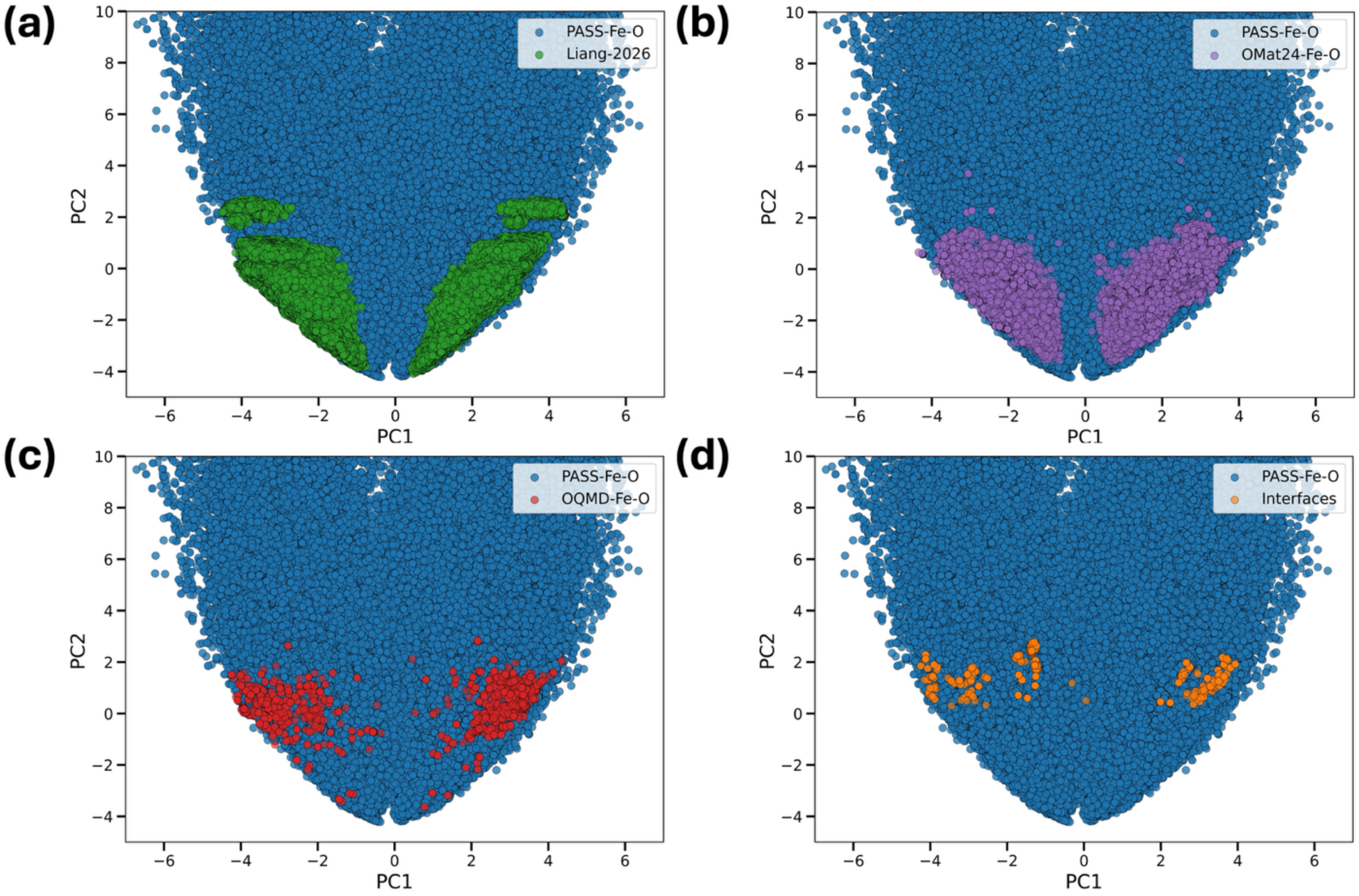

**Fig. S1.** PCA projection of SOAP descriptors for Fe-O structures in PASS dataset, compared with existing Fe-O datasets: (a) Liang-2026[1], (b) OMAT24 Fe-O[2], (c) OQMD Fe-O[3,4] and (d) interfaces structures[5]. Each point represents the LAE of an individual atom.

## Supplementary Figures

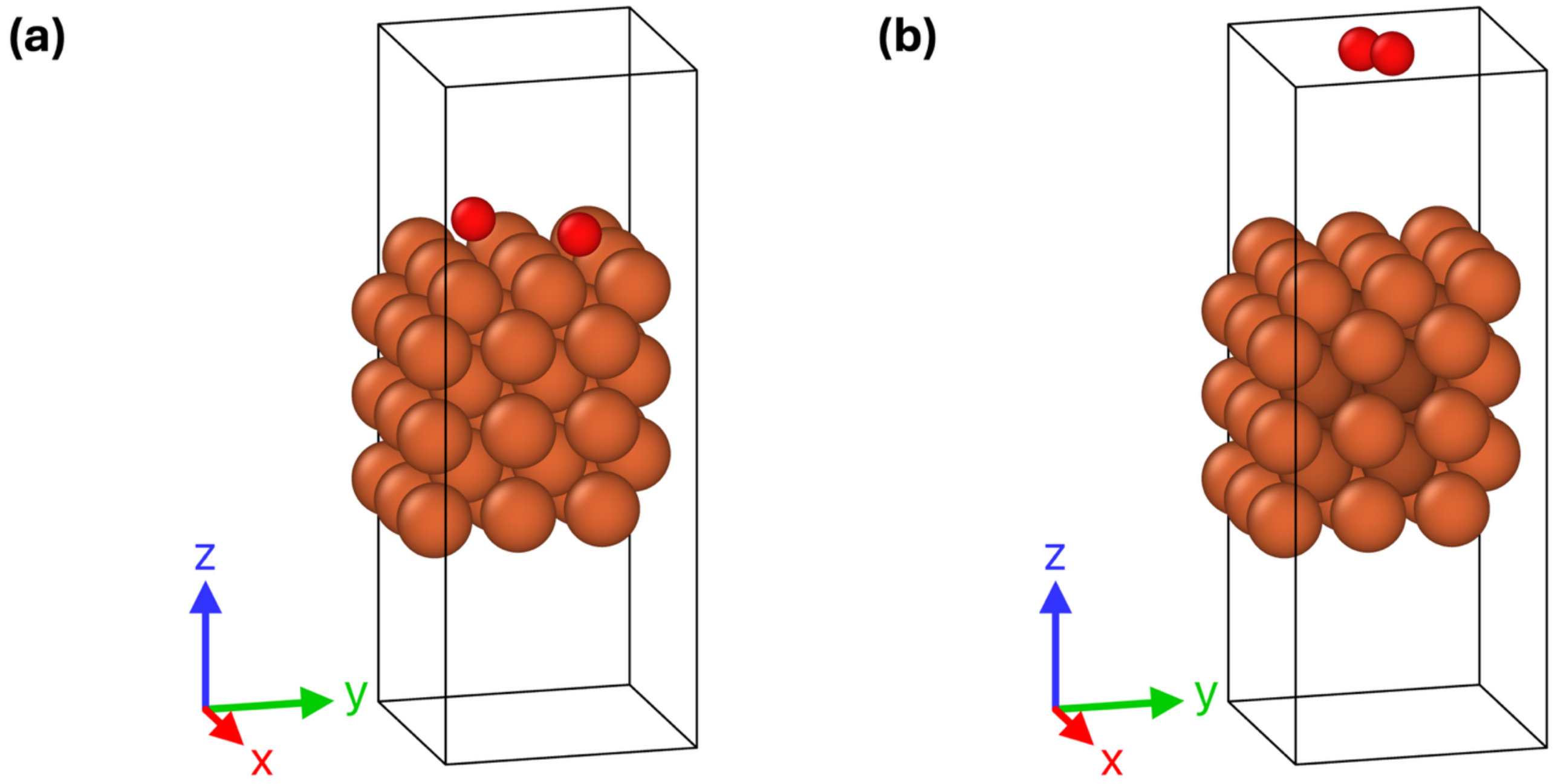

**Fig. S2.** Computational models used to calculate the energy barrier of dissociation adsorption: (a) Initial and (b) final configurations.